\documentclass{statsoc}

\usepackage[a4paper]{geometry}
\usepackage{graphicx}
\usepackage[textwidth=8em,textsize=small]{todonotes}
\usepackage{amsmath}
\usepackage{natbib}
\usepackage{multirow}

\newcommand{\bbetahat}{\mbox{\boldmath $\hat\beta$}}

\title[CADEN]{Cross-validated Risk Scores Adaptive Enrichment (CADEN) Design}
\author[Cherlin and Wason]{Svetlana Cherlin}
\address{Population Health Sciences Institute, Newcastle University, Ridley Building 1, Queen Victoria Road, Newcastle upon Tyne, NE1 7RU, UK.}
\email{svetlana.cherlin@newcastle.ac.uk}
\author[Cherlin and Wason]{James M S Wason}
\address{Population Health Sciences Institute, Newcastle University, Ridley Building 1, Queen Victoria Road, Newcastle upon Tyne, NE1 7RU, UK.\\
MRC Biostatistics Unit, School of Clinical Medicine, University of Cambridge, Cambridge CB2 0SR, UK.}

\begin{document}
\begin{abstract}
We propose a Cross-validated ADaptive ENrichment design (CADEN) in which a trial population is enriched with a subpopulation of patients who are predicted to benefit from the treatment more than an average patient (the sensitive group). This subpopulation is found using a risk score constructed from the baseline (potentially high-dimensional) information about patients. The design incorporates an early stopping rule for futility. Simulation studies are used to assess the properties of CADEN against the original (non-enrichment) cross-validated risk scores (CVRS) design that constructs a risk score at the end of the trial. We show that when there exists a sensitive group of patients, CADEN achieves a higher power and a reduction in the expected sample size, in comparison to the CVRS design. We illustrate the application of the design in two real clinical trials. We conclude that the new design offers improved statistical efficiency in comparison to the existing non-enrichment method, as well as increased benefit to patients. The method has been implemented in an {\tt R} package {\tt caden}.
\end{abstract}

\section{Introduction}

Enrichment clinical trial designs address the issue of investigating treatments that only provide benefit to a subgroup of patients. When the biological pathway of the disease is well understood, predictive biomarkers (biomarkers that can predict the response to treatment) can be identified and the  entry of patient is restricted based on these biomarkers. However, often the mechanism of disease is unclear and there are no known predictive biomarkers. In this case, restricting the entry to a specific subgroup of patients might lead to overlooking other subgroups that also benefit. This leads to loss opportunity for patient-benefit as well as reducing the market for the treatment. 

Adaptive enrichment designs (AEDs) could overcome this limitation by allowing the trial to update the inclusion criteria at the interim analysis. AEDs start by enrolling all the eligible patients and then narrowing the enrolment  of patients to a subgroup predicted to benefit from the experimental treatment, at the interim analysis. Thus, in the second stage, the entry is restricted to the patients who are predicted to benefit from the treatment. For recent review of adaptive enrichment designs, see \cite{baldi_antognini:etal:2023}. Properly constructed adaptive enrichment designs have benefits compared to fixed designs with respect to statistical properties \citep{wang:etal:2009}. 

The subgroup of patients who are predicted to benefit from the treatment is often defined by a single binary or continuous biomarker. \cite {jones:holmgren:2007} considered a design where subjects are classified into a single binary biomarker-positive or biomarker-negative subgroup. At the interim analysis of the preliminary efficacy of the treatment, a decision is made on whether to recruit the overall population or the biomarker-positive group. \cite{parashar:etal:2016} extended the design of \cite{jones:holmgren:2007}  by allowing early efficacy stopping (stopping the trial if the required cumulative response has been achieved) in either the unselected population or the biomarker-positive subgroup at the interim analysis, thus minimising the expected sample size. The design of \cite{tournoux-facon:etal:2011} also allows early stopping for efficacy based on assessing the heterogeneity of response between two subgroups as identified by a single binary biomarker. At the interim analysis, the subgroup in which the treatment effect is uncertain continues to stage 2. \cite{wang:etal:2018} considered an enriched biomarker stratified design in which both biomarker-positive and biomarker-negative patients are selected into the trial with different  probabilities, and a biomarker stratified design in which a (potentially expensive) biomarker is replaced by a low-cost alternative biomarker. When the predictive biomarker is continuous, it could be dichotomised to define two distinct subgroups. However, this requires  pre-specification of an optimal dichotomisation threshold which is often unknown. \cite{wang:etal:2020} and \cite{simon:simon:2017} considered adaptive enrichment designs for a single continuous biomarker with the cut-off threshold being seamlessly determined within the design. 

These designs assume that the sub-populations are characterised by a known single binary or continuous biomarker. However, identifying a single biomarker requires knowledge and biological interpretation of the disease pathway which may not be available. Thus in practice, researchers may opt to utilise a collection of biomarkers. Then subgroups can be defined by this collection. The limitations with current available methods that utilise a collection of biomarkers is that they either: (i) do not address how the composite biomarker should be constructed; (ii) in cases where the method of construction is addressed, are limited to low dimensional covariate data that may assume prior knowledge of predictive biomarkers; or (iii) where the method is tested in high dimensional settings, they are performed on a post hoc basis rather than an enrichment setting. For example, \cite{simon:simon:2013} considered the adaptive enrichment design based on a pre-specified mapping of a space of covariate vector $\bf {x}$ to a \{0,1\}. The mapping is performed according to the probability of response for patients with $\bf {x}$. However, the design assumes that the mapping of the covariate vector $\bf {x}$ is known. \cite{wang:etal:2007} considered an adaptive enrichment design with a subgroup defined by a genomic (composite) biomarker, however the design does not address the issue of constructing the composite biomarker.  \cite{joshi:etal:2020} considered a three-stage adaptive enrichment trial that estimates the best subgroup via a trade-off between the size of the subgroup and the treatment effect in the subgroup. The subgroup was estimated after the first stage and then refined after the second stage based on the accumulated data from the two stages. In this design, a few methods for estimating the subgroup were compared: linear model, overlapping group LASSO \citep{zeng:breheny:2016}, classification and regression trees \citep{brieman:etal:1984}, random forest \citep{brieman:2001}, and support vector machines \citep{cortes:vapnik:1995}. Though simpler (linear model-based) methods were found to perform better than more complex methods for moderate sample sizes and effect sizes, the approach was only evaluated on a low number of biomarkers (four biomarkers). \cite{xu:etal:2020} proposed an adaptive subgroup-identification enrichment design that simultaneously searches for low-dimensional predictive biomarkers, identifies the subgroups with differential treatment effects, and modifies study entry criteria at interim analyses. However, the method assumes knowledge of potential predictive biomarkers which is not always available in practice. The method was also evaluated on a four biomarkers. 

With the recent advances in  multi-omics technologies, an increasingly large numbers of biomarkers (tens of thousands) are becoming available. Several approaches that utilise high-dimensional data have been proposed, such as a family of adaptive signature designs. These include developing a signature from high-dimensional data and then using this signature to identify a subgroup of patients who are the most likely to benefit from the treatment. These approaches perform a post hoc analysis of high-dimensional data rather than adaptively enriching the patient population at the interim analysis. For example, the adaptive signature design \citep{freidlin:simon:2005, freidlin:etal:2010} utilises the information from interaction tests between treatment and (separately) each covariate to develop a genetic signature that can identify a subgroup of patients who benefit from the treatment. In these methods, the development and validation of the signature are performed in a single trial. The adaptive signature design of \cite{zhang:etal:2017} approximated the optimal subgroup on the basis of an arbitrary set of baseline covariates. Several extensions of the adaptive signature design have been proposed, such as adaptive threshold design, generalised adaptive signature design and adaptive signature design with subgroup plots \citep{antoniou:etal:2016}. A compound covariate predictor approach \citep{matsui:2006, matsui:etal:2012, radmacher:etal:2002} and risk scores approach \citep{cherlin:wason:2020} summarise the high-dimensional information into a single score for each patient which is subsequently used for identifying a subgroup of patients who benefit from the treatment. \cite{tian:etal:2014} proposed a method for estimating interaction between covariates and treatment using modified covariates, thus eliminating the need to model the main effect of the covariates. 

We propose a cross-validated risk scores adaptive Enrichment (CADEN) design that  utilises the cross-validated risk scores (CVRS) method of \cite{cherlin:wason:2020} to adaptively enrich the trial with the sub-populations of interest. This design could be used for stage II exploratory trial when an assay or (possibly genetic) signature that identifies sensitive patients is not available. The design employs an interim analysis that explores the sub-population of interest with the CVRS method. Consequently, a decision is made as to whether to enrich the trial with the sub-population of interest, or to proceed with the overall trial population. The design incorporates stopping for futility (in the case where neither the overall treatment effect nor the promising effect in the subgroup is detected). The design is implemented in an open-source {\tt R} package {\tt caden} (Cross-validated ADaptive ENrichment) \citep{github:caden:2021}. 

The remainder of this article is organized as follows. The CVRS section gives a brief recap of the cross-validated risk scores method which is employed in the proposed design. In the CADEN design section, we describe the proposed adaptive enrichment design. In the Simulation Study section explore the operating characteristics of the CADEN designs for various simulation scenarios and compare it to the (non-enrichment) cross-validated risk scores (CVRS) design. In the Real Data Examples section, we illustrate the application of the design to two real clinical trials (the NOAH trial and the PREVAIL trial). Finally, we summarise our conclusions in the Discussion section.

\section{Cross-Validated Risk Scores Design} \label{sec:CVRS} 
In the Cross-Validated Risk Scores method \citep{cherlin:wason:2020}, the (potentially high-dimensional) information about the baseline covariates $L$ is summarised into a single measure (risk score) for each patient $i~(i=1,\dots,n)$ in a two-arm trial (for example, these could be anthropomorphic or genetic data). We assume that the outcome $y_i$ is influenced by a subset of $K$ (unknown) covariates which may be categorial or continuous  (the sensitive covariates) though a generalised linear model. For example, for a binary outcome the model looks as follows:
\begin{equation} \label{eq:model}
\mathrm{logit} (p_i) = \mu + \lambda t_i + \nu_1 x_{i1} + \dots + \nu_K x_{iK} + 
\gamma_1 t_i x_{i1} + \dots + \gamma_K t_i x_{iK},
\end{equation}
where $p_i$ is the probability that the outcome for the $i$th patient is 1, $\mu$ is the intercept; $\lambda$ is the treatment main effect which is independent of the covariates; $t_i$ is the treatment that the $i$th patient receives ($t_i$ = 0 for the control arm and $t_i$ = 1 for the treatment arm); $x_{i1}, \dots, x_{iK}$ are the values for the $K$ unknown sensitive covariates; $\nu_1, \dots, \nu_K$ are the main covariate effects for the $K$ covariates; $\gamma_1, \dots, \gamma_K$ are the treatment-covariate interaction effects for the $K$ covariates.  

Though the identities of the sensitive covariates are unknown, the model assumes that the (positive) interaction between these covariates and the treatment increases the probability of response in a subset of patients (the sensitive group). This leads to the existence of a subgroup with a higher probability of response when treated with the new treatment compared to the control treatment. In the framework of genetics, this might correspond to a subgroup of patients who respond to treatment if they overexpress specific genes.  
For simplicity, we assume that the main treatment effect and all main covariate effects are 0.
This would correspond to $\lambda$ and $\nu_1,\dots,\nu_K$ being 0, whereas $\gamma_1, \dots, \gamma_K$ are non-zero.
	
To construct the risk score for a binary outcome, for each covariate $j$ a single covariate regression is fit:
$\mathrm{logit}(p_i) = \mu + \eta t_i + \nu_j x_{ij} +\beta_j t_i x_{ij}$,
where $\eta$ represents the main treatment effect, $\nu_j$ represents the main effect of the covariate, and $\beta_j$ represent the treatment-covariate interaction effect. 
For each patient $i$, the risk score $RS_i$ is constructed as a sum of the associated covariates $x_{ij}$ weighted by their estimated effects $\hat\beta_j$, \i.e. $\mathrm{RS}_i = \sum_j \hat\beta_j x_{ij}$. The collection of the risk scores $\{RS_i\}$, $i=1,\dots,n$ is then divided into two clusters that correspond to the sensitive and non-sensitive groups of patients, using the $k$-means clustering procedure with $k$ = 2. The $k$-means procedure assigns the risk scores into clusters such that the distances between the risk scores within clusters and the corresponding cluster means (centres) are minimised. Assuming that higher values of the interaction effect increase the probability of response, the cluster with higher cluster mean would correspond to a sensitive group. For constructing the risk scores, the cross-validation procedure is used, in which the model is built using the training subset and the risk scores are constructed for the patients in the test subset. The method is implemented in an {\tt R} package {\tt rapids} \citep{github:rapids:2019}. For full details, including possible filtering of the baseline covariates, see \cite{cherlin:wason:2020}.
	
\section{The CADEN Design}

\begin{figure}[!htbp]
	\centering	
	\includegraphics[scale=0.7]{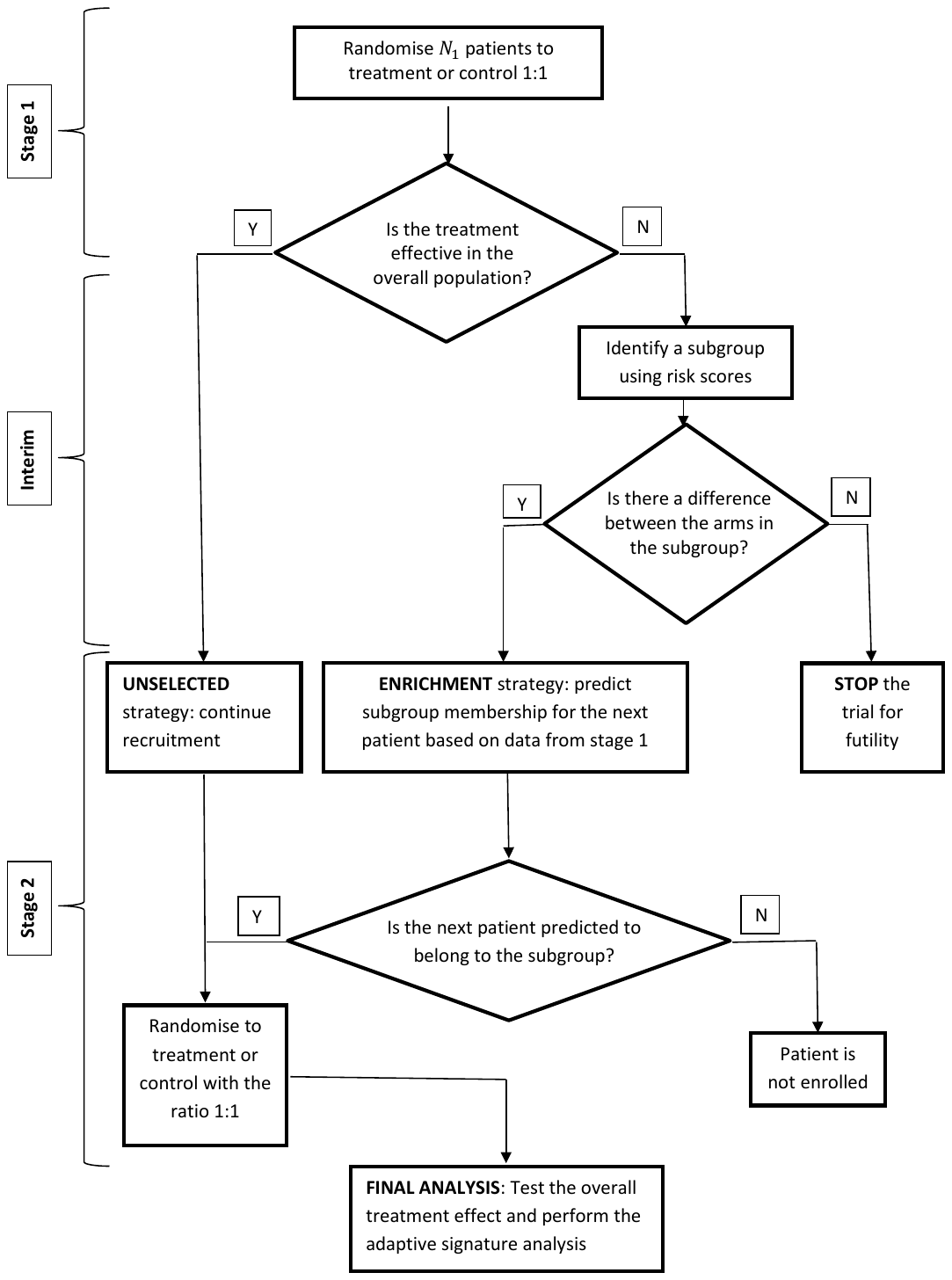}
	\caption{The flowchart of the design.}
	\label{fig:flowchart}
\end{figure}

The CADEN design consists of two stages, as illustrated in Figure \ref{fig:flowchart}. In stage 1, $N_1$ patients are enrolled and  randomised to an experimental arm or a control arm. Without loss of generality, we assume	randomisation is equal to each treatment.

At the end of stage 1, an interim analysis is performed to test the efficacy of the treatment compared to the control in the overall population, at a pre-specified  significance level $\alpha_{1}$. If the test in the overall population meets the $\alpha_{1}$ significance level then the trial proceeds into stage 2 by enrolling all eligible patients (the ``unselected" strategy, see Section \ref{sec:unselected}). If the test in the  overall population does not meet the $\alpha_{1}$ significance level then the trial  proceeds by identifying a subgroup of patients that have a higher probability of response than the overall trial population (the sensitive group). Here, a sensitive group is identified by applying the cross-validated risk scores analysis (see Section \ref{sec:CVRS}).  If there is a significant difference in the response rates between the arms within the sensitive group at a pre-specified significance level $\alpha_{2}$, the trial proceeds according to the ``enrichment" strategy as described in Section  \ref{sec:enrichment}, otherwise the ``stop" strategy is employed (see Section  \ref{sec:stop}). For the final analysis of the ``enrichment" strategy, the $\alpha$-splitting approach is used to adjust for multiple testing. 

\subsection{``Unselected" strategy} \label{sec:unselected}

If at the interim analysis the treatment is found to be effective in the overall population at the significance level $\alpha_{1}$  (we use $\alpha_{1} = 0.05$), then the  recruitment continues from the overall (unselected) eligible population of patients, until the planned sample size $N$ is reached (there is no stopping for efficacy). At the final analysis, the difference between the treatment arms is evaluated in the overall population of patients, using the data from both stages. Additionally, a post hoc cross-validated risk scores analysis is performed (see Section \ref{sec:CVRS}).

\subsection{``Stop" strategy} \label{sec:stop}

If at the interim analysis the treatment is not found to be effective in the overall population at the significance level $\alpha_{1}$, and is also not found to be promising in the sensitive subgroup at the significance level $\alpha_{2}$ then the trial stops for futility.

\subsection{``Enrichment" strategy}\label{sec:enrichment}

It might happen that at the interim analysis the treatment is not found to be effective in the overall population at the significance level $\alpha_1$ but there is a significant difference between the arms in the subgroup at the significance level $\alpha_2$. (To explore the operational characteristics of the design, we use three difference significant level of $\alpha_2$ in the simulation study ($\alpha_2$ = 0.05, 0.1, 0.2)). In this case, the trial is enriched with the patients who are predicted to belong to the sensitive group, i.e. only patients who are predicted to belong to the sensitive subgroup are recruited for stage 2. 
To predict whether a patient would belong to the sensitive group, we compute their risk score using the coefficients $\bbetahat$ obtained by applying the adaptive signature analysis to stage 1 data, as described in Section \ref{sec:CVRS}.  Here, the cross-validation technique is not required, because the $\bbetahat$ coefficients would be applied to new data. Additionally, we compute the means of the sensitive and the non-sensitive clusters of the risk scores from stage 1 data ($m_{S}$ and $m_{N}$, respectively). For every new patient, the following steps are performed:
\begin{itemize}
	\item [(i)] A risk score $r$ is computed $r = \bf {x}\bbetahat^T$ where  $\bf x$ is covariate data for a new patient, and $\bbetahat$ are coefficients obtained from applying adaptive signature analysis to stage 1 data;
	\item [(ii)] Predicted sensitivity status is computed according to the distance between $r$ and one of the cluster means $m_{S}$ and $m_{N}$ as follows:
	the new patient is predicted to belong to the sensitive group if $m_S - r < m_{N} - r$, i.e. if the risk score for this patient is close to the mean of the sensitive cluster;
	\item [(iii)] The patient is randomised to either the  control or treatment arm with the ratio 1:1 if they are predicted to belong to the sensitive subgroup, otherwise they will not be eligible to take part in the trial.
\end{itemize}
Steps (i) - (iii) are performed until the planned sample size $N$ is reached. The final analysis is performed using the sensitive patients from stage 1 and all the patients from stage 2 because they are all predicted to be sensitive. To generalise the findings, the coefficients $\bbetahat$ and  the cluster means $m_{S}$ and $m_{N}$ could be used in a future phase III confirmatory trial, to identify patients who would benefit from the new treatment. 

\subsection {Hypotheses testing}\label{sec:hypotheses}

In the CADEN design we consider a dual-composite null hypothesis $H_C$ \citep{wang:hung:2013} that states that there is no difference between the response rates in the treatment arms in the overall trial population and also no difference between the response rates in the treatment arms within the sensitive group (the sensitive group is obtained by applying the adaptive signature analysis as described in Section \ref{sec:CVRS}).
The null hypothesis $H_C$ can be represented as an intersect of the hypothesis $H_O$ and $H_S$, i.e.  
$H_C = H_O \cap H_S$. In this equation, the $H_O$ represents the hypothesis that there is no difference between the response rates in the treatment arms in the overall trial population. For a binary outcome, this could be evaluated using a test based on a normal approximation for the difference of two proportions ({\tt prop.test}  function in {\tt R}, two-sided). $H_S$ represents the hypothesis that there is no difference between the response rates in the treatment arms within the sensitive group. For a binary outcome, this could be examined using Fisher's exact test (two-sided). The $p$-value for testing the $H_S$ is computed using the Stouffer's combination method of the $p$-value from testing the difference between the arms in the sensitive group in Stage 1, and the $p$-value from testing the difference between the arms in Stage 2 (here, all patients belong to the sensitive group by design).

\section{Simulation Study} 

\subsection{Simulation settings}

We consider three simulation settings.
In setting (i), we assume that there is no overall treatment effect ($\lambda = 0$) and also no main effect of the covariates ($\alpha_1, \dots, \alpha_K = 0$). However, there exists a sensitive group of patients for whom the interaction effect between treatment and sensitive covariates increases the probability of the response to treatment. Within this setting, we consider scenarios with different response rates in the sensitive group, and also different sample sizes. Setting (ii) corresponds to a null scenario in which the sensitive group does not exists. Within this setting, we consider a scenario where the response rate is the same for the entire trial population, and a scenario where there is an overall treatment effect ($\lambda \neq 0$).  In setting (iii) we consider the situation where the treatment appears to be harmful for some patients. In this setting, we analyse two scenarios. In both scenarios, there is a subgroup of patients with a negative treatment-covariate interaction effect for some covariates, so that the response rate for this subgroup is lower than that for the control. In the second scenario, there is also a sensitive group, i.e a subgroup of patients with positive treatment-covariate interaction effect for the sensitive covariates. In both scenarios, we assume that there is no overall treatment effect and also no main effect of the covariates. See Appendix and Table \ref{tab:Table1} for details of how the data were simulated. The parameters were chosen in accord with the previously published adaptive signature designs \citep{freidlin:simon:2005, freidlin:etal:2010}.

\begin{table} 
	\caption{\label{tab:Table1} Response rate, prevalences and the sample sizes used in simulation study. The ``$RR_1$" column represents the response rate in the sensitive group on the treatment arm. The ``$RR_2$" column represents the  response rate in the non-sensitive group on the treatment arm. The ``$RR_0$" column represents the response rate in all patients on the control arm. The ``$RR_3$" column represents the response rate in the subgroup of patients who experience a harmful effect on the treatment arm. The ``Prevalence" column represents the prevalence of the sensitive group and/or a group with a harmful effect on the treatment arm.}
	\centering
	\begin{tabular}{ccccccccc} \hline
		{Setting} & 
		{Table} & 
		{Scenario}  & $RR_1$ & $RR_2$ & $RR_0$ & $RR_3$ & Prevalence & Sample Size\\ \hline
		i &	 1  & - & 0.5  & 0.25     & 0.25 & - & 0.1 & 400 \\
		i &	 1  & - & 0.6  & 0.25     & 0.25 & - & 0.1 & 400 \\
		i &	 1  & - & 0.7  & 0.25     & 0.25 & - & 0.1 & 400 \\
		i &	 1  & - & 0.5  & 0.25     & 0.25 & - & 0.1 & 1000 \\
		i &	 1  & - & 0.6  & 0.25     & 0.25 & - & 0.1 & 1000 \\
		i &	 1  & - & 0.7  & 0.25     & 0.25 & - & 0.1 & 1000 \\
		i &	3 &  A  & 0.6     & 0.25     & 0.25 & - & 0.2 & 400  \\
		ii & 2&  A  & 0.25    & 0.25     & 0.25 & - & - & 400  \\
		ii & 2 & B  & 0.35     & 0.35     & 0.25 & - & - & 400  \\
		iii & 3 & B  & 0.25   & 0.25     & 0.25 & 0.1 & 0.2 & 1000  \\
		iii & 3 & C  & 0.4     & 0.25     & 0.25 & 0.1 & 0.2 & 100 \\ \hline
	\end{tabular}
\end{table}

To the best of our knowledge, there are no adaptive enrichment designs that utilise high-dimensional data. Therefore, to undertake the comparison to other enrichment methods, we compared the CADEN design with the benchmark design where the sensitive group is determined by a (known) single binary biomarker. 	We regard this as a hypothetical upper bound on how the CADEN approach could perform as we are assuming there is actually no known true biomarker. We also considered designs where (i) only 80\% of biomarker-positive patients belong to the sensitive group (this corresponds to a single biomarker being 100\% sensitive and 80\% specific, when 20\% of the biomarker-negative samples are mistakenly classified as biomarker-positive) and (ii) all of the biomarker-positive patients belong to the sensitive group, however, the biomarker has 95\% sensitivity and 95\% specificity, i.e. 5\% of the of the biomarker-positive samples are mistakenly classified as biomarker-negative, and 5\% of the biomarker-negative samples are mistakenly classified as biomarker-positive.

\subsection{Operating characteristics}

We perform 1,000 runs for each simulation scenario and compute three types of statistical power. 
The first power hereafter referred to as $PWR_{O}$ is the power to reject $H_{O}$ when $H_O$ is not true, that is the power to detect a difference in response rates between the treatment arms at the significance level $\alpha_O$ when the difference is indeed present. It is computed as a proportion of simulation runs with $p < \alpha_O$.

The second power hereafter referred to as $PWR_{S}$ is the power to reject $H_{S}$ when $H_{S}$ is not true, that is the power to detect a difference in response rates between the treatment arms in the sensitive group at the significance level $\alpha_S$ for the ``unselected" strategy, or at the significance level $\alpha$ for the ``enrichment" strategy, where $\alpha = \alpha_O + \alpha_S$. $PWR_S$ is computed as a proportion of simulation runs where the sensitive subgroup hypothesis is tested and rejected. 

The third power, hereafter referred to as $PWR_{C}$ is the power to reject the dual-composite hypothesis $H_C$ when $H_C$ is not true, that is the power to detect either a difference between the arms in the overall population or in the sensitive group. In other words, $PWR_C$ is the power to reject either null hypothesis $H_O$ or $H_S$. It could be interpreted as the power to conclude that the trial is positive. $PWR_C$ is computed as the proportion of simulations in which one of $H_O$ or $H_S$ is rejected. We note that for the null scenario, $PWR_{C}$ would correspond to FWER which is a chance of rejecting any true null hypothesis. Additionally, for each simulation scenario, we estimate the sensitivity and specificity of the design to predict the membership in the sensitive group of patients, and the expected sample size. Sensitivity and specificity of identifying the sensitive group were computed as the probability to correctly identify patients as sensitive and the probability to correctly identify patients as non-sensitive, respectfully. 
 The expected sample size, $N_{exp}$, is computed as $N_{exp} = N_1 \times \eta_{s} + (N_1 + N_2) \times (1-\eta_{s})$, where $\eta_{s}$ is the proportion of the simulation runs that follow the ``stop" strategy, and $N_1$ and $N_2$ are sample sizes for stage 1 and 2, respectively.

\subsection{Simulation results}

Table \ref{tab:Table2} presents the results for the simulation setting (i) in which there exists a sensitive group of patients. Here, we analyse scenarios with different sample sizes (400 and 1000) and different response rates for the sensitive group on treatment (50\%, 60\% and 70\%). For each case, we investigate three thresholds for $\alpha_2$ (0.05, 0.1 and 0.2), the significance level for identifying a promising treatment effect in the sensitive group at the interim analysis.  The sensitivity and specificity of identifying the sensitive group is high, reaching values of larger than 0.9 in most cases. As expected, the power is larger for a larger sample size and for a higher response rate in the sensitive group. For the same sample size and the same response rate in the sensitive group, the power increases as $\alpha_2$ increases because increasing $\alpha_2$ increases the probability of the design to follow the ``enrichment" strategy which would be the true strategy in this scenario. For example, with a sample size 1000 and a 50\% response rate in the sensitive group, the power to reject the dual-composite null hypothesis ($PWR_C$) is 0.365, 0.47 and 0.58 for $\alpha_2$ = 0.05, 0.1 and 0.2, respectively, while for the 60\% response rate in the sensitive group $PWR_C$ is 0.64, 0.72 and 0.82. 

\begin{table} 
	\caption{\label{tab:Table2} 
		Operating characteristics of the adaptive risk scores enrichment (CADEN) design. 
		The response rate on the control arm, and for non-sensitive patients on the treatment arm is 25\%. 
		The prevalence of the sensitive patients is 10\%. 
		The sample sizes for each stage are equal to a half of the overall sample size. 
		Ten covariates out of 100 are sensitive. 
		The results, as based on 1000 simulations, are presented for 
		thresholds $\alpha_{2}$ = 0.05/0.1/0.2 for the treatment effect in the sensitive 
		group at the interim analysis. Sensitivity and specificity for stage 1 are based on all the simulations, 
		sensitivity and specificity for stage 2 are based on the simulations 
		with the ``enrichment" strategy only.}
	\centering
	\begin{tabular} {lcccccc} \hline
		& \multicolumn{3}{c}{Response rate in the sensitive group}  \\ \cline{2-4}  
		Operating \\ characteristics & 50\% & 60\% & 70\% \\ \hline 
		& \multicolumn{2}{c}{\centering Sample size = 400 } \\ \hline
		$PWR_O$ &0.02/0.0/0.02 & 0.02/0.03/0.02 & 0.03/0.05/0.04 \\
		$PWR_S$ & 0.07/0.16/0.22 & 0.17/0.26/0.38 & 0.29/0.41/0.54 \\
		$PWR_C$ & 0.09/0.18/0.24 & 0.18/0.27/0.39 & 0.3/0.42/0.55 \\
		Sensitivity stage 1 & 0.89/0.89/0.89 & 0.96/0.96/0.95 & 0.99/0.98/0.99 \\  
		Specificity stage 1 & 0.75/0.75/0.75 & 0.83/0.83/0.83 & 0.88/0.88/0.88 \\
		Sensitivity stage 2 & 0.92/0.94/0.91 &0.99/0.98/0.98 & 0.99/0.99/0.99 \\  
		Specificity stage 2 & 0.9/0.92/0.91 & 0.98/0.97/0.97 & 1.00/1.00/0.99 \\
		Unselected strategy, \%  &5.6/5.6/5.6 & 4.2/4.2/5.4 &  6.9/6.9/6.9 \\		
		Enrichment strategy, \% &7.4/17.3/26.3& 15.5/24.6/38.3  & 24.2/36.1/49.4 \\
		Stop strategy, \%  & 87/77.1/68.1 & 80.3/71.2/57.5 &   68.9/57/43.7\\
		Expected sample size & 226/246/264 & 240/258/285 &  263/286/313 \\ \hline
		& \multicolumn{2}{c}{\centering Sample size = 1000 } \\
		\hline 
		$PWR_O$ & 0.04/0.04/0.04 & 0.08/0.08/0.07 & 0.13/0.13/0.14 \\
		$PWR_S$ & 0.35/0.46/0.56 & 0.64/0.72/0.82 & 0.87/0.91/0.95 \\
		$PWR_C$   & 0.36/0.47/0.58 & 0.64/0.72/0.82 & 0.87/0.91/0.96 \\
		Sensitivity stage 1 & 0.97/0.97/0.9 7& 0.99/0.99/0.99 & 1.00/1.00/1.00 \\
		Specificity stage 1 & 0.92/0.92/0.92 & 0.97/0.97/0.97 & 0.99/0.99/0.99 \\
		Sensitivity stage 2  & 0.99/0.99/0.99 & 1.00/0.99/0.99 & 1.00/1.00/1.00 \\
		Specificity stage 2  & 1.00/0.99/0.99 & 1.00/1.00/1.00/ & 1.00/1.00/1.00 \\
		Unselected strategy, \%& 7.4/7.4/7.4 & 12.4/12.4/12.4 & 17.6/17.6/17.6 \\
		Enrichment strategy, \% & 29.9/41.5/52.4 & 60.8/70.5/76.5 & 68.2/73.3/78.2 \\
		Stop strategy, \%   & 62.7/51.1/40.2 & 35.4/26.8/17.1 & 14.2/9.1/4.2 \\
		Expected sample size & 687/745/799 & 823/866/915 & 929/955/979 \\
		\hline
	\end{tabular}
\end{table}

Table \ref{tab:Table3} presents the results for simulation setting (ii) in which there is no sensitive group. In this setting, we consider two scenarios: scenario A (a null scenario), where the response rate for everyone is 25\%, and scenario B where everyone benefits equally from the treatment (the response rates are 25\% and 35\% on the control and the treatment arms, respectively). For scenario A, the power for rejecting the dual-composite null hypothesis ($P_C$) is less than 0.05 which means that the type I error rate is well controlled. For scenario B, $P_C$ is 0.3 - 0.35 due to the fact that there is a difference between the treatment groups in the overall population. The sensitivity and the specificity of identifying the sensitive group is around 0.5 for both scenarios which reflects the fact that there is no true underlying sensitive group. Additionally, the reduction in the expected sample sizes ranges between 40\%-46\% for scenario A and between 26\%-33\% for scenario B, in comparison to the maximum sample size.

\begin{table} 
	\caption{\label{tab:Table3} Operating characteristics of the adaptive risk scores enrichment (CADEN) design. The response rate on the control arm, and for non-sensitive patients on the treatment arm is 25\%. Ten covariates out of 100 are sensitive. The results are based on 5000 simulations. Sample sizes are 200 for stage 1 and 200 for stage 2. The results are presented for thresholds $\alpha_2$ = 0.05/0.1/0.2 for the treatment effect in the sensitive group at the interim analysis. Sensitivity and specificity for stage 1 are based on all the simulations, sensitivity and specificity for stage 2 are based on the simulations with the ``enrichment" strategy only. Scenario A: The response rate for everyone on both arms is 25\% (null scenario). Scenario B: The response rate for everyone on the treatment arm is 35\%.}
	\centering
	\begin{tabular} {lcc}\hline
		Operating  characteristics & Scenario A & Scenario B \\ \hline 
		$PWR_O$ & 0.01/0.01/0.01 & 0.25/0.25/0.25\\ 
		$PWR_S$ & 0.01/0.02/0.03 & 0.11/0.13/0.17\\
		$PWR_C$ & 0.02/0.03/0.04 & 0.3/0.32/0.35\\
		Sensitivity stage 1 & 0.48/0.49/0.49 & 0.55/0.55/0.55\\
		Specificity stage 1 & 0.50/0.50/0.50 & 0.53/0.53/0.53\\
		Sensitivity stage 2 & 0.47/0.47/0.50 & 0.52/0.52/0.52 \\
		Specificity stage 2 & 0.55/0.55/0.57 & 0.59/0.58/0.58\\
		Unselected strategy, \%& 3.2/3.3/3.3 & 29.2/29.2/29.2\\
		Enrichment strategy, \%& 3.3/7.8/16.2 & 6.1/10.3/18.6\\
		Stop strategy, \%      & 93.5/88.9/80.5 & 64.7/60.5/52.2\\
		Expected sample size & 214/223/239 & 271/279/296 \\ \hline
	\end{tabular}
\end{table}

We also compare the performance of the CADEN design with the CVRS design for three different  scenarios (Table \ref{tab:Table4}). In scenario A, there exists a sensitive group of patients who benefit from the treatment more than everyone else. In this scenario, the response rate for the sensitive group on treatment is 60\%, the prevalence of the sensitive group is 20\%, and the response rate for the non-sensitive patients on treatment and for everyone in the control arm is 25\%. The comparison of the power to reject $H_S$ ($PWR_S$), as well as the power to reject $H_C$ ($PWR_C$) between the CADEN and the CVRS designs depend on the threshold for the difference in the sensitive group ($alpha_2$). For CADEN, a reduction in expected sample size is achieved due to the adaptive nature of the design (specifically, due to stopping for futility). For comparison, we computed the operating characteristics for the CVRS using the expected sample size achieved with CADEN (351 patients) and show that for the same expected sample size, CADEN achieves higher power than CVRS. The power to reject $H_O$ is smaller for the CADEN design because the design less commonly proceeds according to the ``unselected" strategy. In Scenario B, the treatment appears to be harmful rather than beneficial for the sensitive group that comprises 20\% of patients (the response rate for the sensitive group on treatment is 10\%, while the response rate for everyone else is 25\%). Low values of the sensitivity and specificity reflect the fact that the ``non-sensitive group" is actually the sensitive group that experiences higher benefit from the treatment (although there is no benefit compared to control). The expected required sample size and the power for CADEN is much smaller than that for the CVRS which is desirable for this scenario. The power to conclude that the trial is positive which represents the type 1 error rate in this scenario ($PWR_C$) is lower for CADEN in comparison to CVRS. Also, the expected sample size is smaller. In scenario C, we assume that there is a sensitive group of patients with a prevalence of 20\% which benefits from the treatment (the response rate is 40\%). In addition to the sensitive group, there is another group of patients with a prevalence of 20\% who experience harmful effects to treatment (the response rate is 10\%). In this scenario, CADEN shows an exceptional performance, effectively stopping at the interim analysis thus achieving a 50\% reduction in sample size. A high sensitivity and a low specificity for both the CADEN and the CVRS designs reflect the fact that the underlying design assumes the existence of two groups (sensitive and non-sensitive) while in reality there are three groups (sensitive, non-sensitive and a group with a harmful effect). In this situation the true sensitive group is well identified, while the non-sensitive group is not.

\begin{table} 
	\caption{\label{tab:Table4} Comparison of the cross-validated adaptive enrichment (CADEN) design with the CVRS design. The response rate on the control arm is 25\%.  Ten covariates out of 100 are sensitive. For CADEN, the results as presented for thresholds $\alpha_2$ = 0.05/0.1/0.2 for the treatment effect in the sensitive group at the interim analysis. Sensitivity and specificity for stage 1 are based on all the simulations, sensitivity and specificity for stage 2 are based on the simulations with the ``enrichment" strategy only. 
		Scenario A: The response rate for the sensitive group on the treatment arm is 60\%, the response rate for non-sensitive patients on the treatment arm is 25\%. The prevalence of the sensitive patients is 20\%. Sample size is 400 for CVRS, 200 for each stage of CADEN.    
		Scenario B: The treatment appears to be harmful for 20\% patients (response rate 10\%), the response rate for the rest of the patients on the treatment arm is 25\%.  Sample size is 1000 for CVRS, 500 for each stage of CADEN. 
		Scenario C: The response rate for the sensitive group on the treatment arm is 40\%, the response rate for non-sensitive patients on the treatment arm is 25\%. The prevalence of the sensitive patients is 20\%. There are 20\% of patients with a response rate of 10\% on the treatment arm.  Sample size is 1000 for CVRS, 500 for each stage of CADEN.}
	\centering
	\begin{tabular} {l|cc|cc|cc}	\hline
		Operating  & 
		\multicolumn{2}{c|}{Scenario A} &
		\multicolumn{2}{c|}{Scenario B} &
		\multicolumn{2}{c}{Scenario C} \\
		\cline{2-7} characteristics & 
		CADEN & CVRS & CADEN & CVRS & CADEN & CVRS \\ \hline
		$PWR_O$  & 0.12/0.1/0.11 & 0.26/0.22 & 0.05/0.05/0.06 & 0.15 & 0/0/0 & 0.04\\ 
		$PWR_S$ & 0.53/0.62/0.73 & 0.68/0.58 & 0.01/0.01/0.02 & 0.002 & 0/0/0 & 0.13\\ 
		$PWR_C$ & 0.54/0.63/0.74 & 0.77/0.67 & 0.06/0.06/0.07 & 0.15 & 0/0/0 & 0.16\\ 
		Sensitivity stage 1 & 0.97/0.97/0.97 & 0.99/0.99 & 0.05/0.05/0.05 & 0.01 & 0.99/0.99/0.99 & 1.00\\  
		Specificity stage 1 & 0.94/0.94/0.94 & 0.99/0.99 & 0.08/0.08/0.09 & 0.004 & 0.62/0.62/0.62 & 0.63\\  
		Sensitivity stage 2 & 0.99/0.98/0.98 & - & 0.06/0.06/0.06& - & - & -\\  
		Specificity stage 2 & 1.00/0.99/0.99 & - & 0.1/0.09/0.08 & - & - & -\\
		Expected sample size & 310/329/351 & 400/351 & 551/564/602 & 1000 & 500/500/500 & 1000\\ \hline  
	\end{tabular}
\end{table}

Table \ref{tab:Table5} shows the results for the comparison of the CADEN design with the benchmark scenario of a known single biomarker, and with two additional scenarios where a single biomarker is not 100\% sensitive/specific. As expected, the CADEN design performs worse than the benchmark scenario in which the biomarker is perfectly classified (CADEN-BM1 and CADEN-BM2). However, it performs almost comparably to a more realistic CADEN-BM3 scenario  in which the biomarker is not 100\% sensitive/specific.

\begin{table} 
	\caption{\label{tab:Table5} A comparison of the CADEN method (the ``CADEN" column) to different methods in which the sensitive group is determined by a single known binary biomarker (the ``CADEN-BM1", the ``CADEN-BM2" and the ``CADEN-BM3" columns).  The ``CADEN" column corresponds to one of the investigated scenarios presented in Table \ref{tab:Table1}. The ``CADEN-BM1" column corresponds to a scenario in which the biomarker is perfectly classified, and all of the biomarker-positive patients belong to the sensitive group. The ``CADEN-BM2" column represents a scenario where a biomarker is perfectly classified, however only 80\% of the biomarker-positive patients belong to the sensitive group. The ``CADEN-BM3" column corresponds to a scenario where all of the biomarker-positive patients belong to the sensitive group, however the biomarker has a 95\% sensitivity and a 95\% specificity, i.e. 5\% of the of the biomarker-positive patients are mistakenly classified as biomarker-negative, and 5\% of the biomarker-negative patients are mistakenly classified as biomarker-positive. In all scenarios, the samples size 400, the response rate for the treatment arm in the sensitive group is 0.7, the prevalence of the sensitive group is 0.1, the response rate on the control arm and on the treatment arm for the non-sensitive group is 0.25. 
		Sensitivity and specificity for stage 1 are based on all the simulations, sensitivity and specificity for stage 2 are based on the simulations with the ``enrichment" strategy only. }
	\centering
	\begin{tabular} {lcccc} \hline  
		Operating \\ characteristics &CADEN & CADEN-BM1 & CADEN-BM2 & CADEN-BM3 \\ \hline 
		$PWR_O$ & 0.03/0.05/0.04 & 0.05/0.05/0.03 & 0.04/0.04/0.04 & 0.03/0.04/0.04\\
		$PWR_S$ & 0.29/0.41/0.54 & 0.37/0.52/0.68 & 0.36/0.52/0.69 & 0.32/0.41/0.55\\
		$PWR_C$ & 0.3/0.42/0.55 & 0.38/0.53/0.69 & 0.38/0.52/0.71 & 0.32/0.42/0.56\\
		Sensitivity stage 1   & 0.99/0.98/0.99 & 1.00/1.00/1.00 & 1.00/1.00/1.00 & 0.95/0.95/0.95\\  
		Specificity stage 1   & 0.88/0.88/0.88 & 1.00/1.00/1.00 & 0.98/0.98/0.98 & 0.95/0.95/0.95\\
		Sensitivity stage 2   & 0.99/0.99/0.99 & 1.00/1.00/1.00 & 1.00/1.00/1.00 & 0.95/0.95/0.95\\ 
		Specificity stage 2   & 1.00/1.00/0.99 & 1.00/1.00/1.00 & 0.98/0.98/0.98 & 0.95/0.95/0.95\\
		Unselected strategy, \%  & 6.9/6.9/6.9 & 6.9/6.9/6.9 & 6.9/6.9/6.9 & 6.9/6.9/6.9\\		
		Enrichment strategy, \% & 24.2/36.1/49.4 & 31.8/47.2/64 & 31.8/47.2/64.0 & 26.5/35.8/50.6\\
		Stop strategy, \%    & 68.9/57/43.7 & 61.3/45.9/29.1 & 61.3/45.9/29.1 & 66.6/57.3/42.5\\
		Expected sample size  & 263/286/313  & 278/309/342 & 278/309/342 & 267/286/315\\ \hline
	\end{tabular}
\end{table}

\section{Real Data Examples}

\subsection{The NOAH trial}

We applied our approach to a publicly available dataset with high-dimensional
gene expression biomarker data from the randomised controlled trial NOAH \citep{gianni:etal:2010}. The trial enrolled 235 patients with HER2-positive locally advanced breast cancer, 117 of whom treated with neoadjuvant chemotherapy with trastuzumab followed by  adjuvant trastuzumab, and the rest (118 patients) treated with neoadjuvant chemotherapy alone. The primary outcomes of the trial were pathological complete response and event-free survival. We used the pathological complete response as an outcome because it was available in the publicly available dataset. We used a binary outcome with achieving pathological complete response having the value of one and residual disease having the value of zero.

Gene expression data from core biopsies that were prospectively collected before treatment,  are available for 114 HER2-positive patients from the National Center for Biotechnology Information (NCBI) website (GSE50948), of whom 111 had the response data recorded. Of the 111 patients, 60 patients were in the treatment group, and the remaining 51 patients were in the control group. From a total of 54,675 genes, we restricted our analysis to the 5,000 probes with the highest variability, to adjust for the relatively small sample size \citep{freidlin:etal:2010}. For the same consideration, we analysed the 111 patients as if they belonged to a stage 1 of a hypothetical two-stage trial. For the hypothetical stage 2, we repeatedly sampled 111 patients with replacement from the original pool of patients. Analysis of stage 1 resulted in the ``unselected" strategy decision ($p$ = 0.03 for the difference between the arms in the overall trial population). To sample patients for stage 2, we performed 1,000 runs with different seeds for random number generator, resulting in different set of patients for every run. The power for the final analysis was computed as the proportion of  runs with significant $p$-values. The final analysis identified a significant difference between the treatment arms in the overall population ($PWR_O$ = 0.944) and a significant treatment effect in the sensitive group of patients ($PWR_S$ = 1) which resulted in high power to reject the dual-composite null hypothesis ($PWR_C$ = 1). To compare the CADEN design with the CVRS design, we applied the CVRS method  to 1,000 data sets. Each data set consisted of 222 patients, of whom 111 were the original patients, and another 111 were repeatedly sampled with replacement from the original pool of patients for the stage 2 of the CADEN analysis. This resulted in the same power as that for the CADEN analysis. 

\subsection{The PREVAIL trail}
We applied our method to a publicly available dataset with high-dimensional
gene expression biomarker data from the PREVAIL study \citep{muscedere:etal:2018} 
which was a phase II/III randomized controlled multi-centred trial examining the use of lactoferrin to prevent nosocomial infections in critically ill patients undergoing mechanical ventilation. Out of 212 patients who formed the analysis cohort, 107 received lactoferrin and 105 received placebo. The trial concluded that lactoferrin improved neither the primary outcome of
antibiotic-free days, nor any of the secondary outcomes which were ICU length of stay, hospital length of stay and hospital mortality.

Gene expression data that was generated from a consecutive subset of patients at the lead study site at various time points during the ICU stay, are available from the NCBI website (GSE118657). We analysed 61 patients who had gene expression data generated for day 1 of the ICU stay. We used ICU mortality as the primary outcome because it was reported in the publicly available dataset. Out of 61 patients who had the genetic data and the primary outcome reported, 32 were in the treatment (lactoferrin) group and 29 were in the control (placebo) group. Similarly to the NOAH trial, we analysed 5,000 genes with the highest variability (out of 49,495), and we analysed all the patients as if they belonged to a stage 1 of a hypothetical two-stage trial. 

Analysis of stage 1 resulted in a non-significant difference  between the arms in the overall trial population ($p$ = 0.4), therefore we evaluated the treatment effect in the sensitive group. Because the sensitive group was obtained using the same data (by the means of cross-validation), we used a permutation method (with 5,000 permutations) to obtain a valid $p$-value \citep{simon:etal:2004}. After randomly permuting the treatment labels, we obtained the sensitive group using cross-validation and computed the $p$-value for the difference between the arms in the sensitive group. The process was repeated 5,000 times.
We computed the permutation-based $p$-value for the treatment effect in the sensitive group as
$\frac{1+\mathrm{number~of~permuted~data~sets~with~} p < p_0}{1+\mathrm{number~of~permuted~data~sets}}$
where  $p$ and $p_0$ are $p$-values for the difference between the arms in the sensitive group in the permuted and the original (non-permuted) data, respectively.
The $p$-value for the original (non-permuted) data, $p_0$, was computed as a mean of 1,000 $p$-values obtained using different seeds for identifying the sensitive group via cross-validation. This was done to take into account the variability of the data due to small sample size. The final permutation-based $p$-value was 0.71 which did not show a promising treatment effect, thus leading to the ``stop" strategy decision.

\section{Discussion}

Adaptive enrichment designs have the potential to improve the efficiency of clinical trials by enriching the trial population with a subgroup of patients that could benefit from the treatment more than average (the sensitive group). Several statistical approaches have been proposed so far in the literature for identifying the sensitive group using biomarker data. Here, we presented an adaptive enrichment design that identifies the sensitive group based on (potentially high-dimensional) baseline data. Specifically, the design considers enriching the recruitment with patients who are predicted to benefit from the treatment, based on their baseline  covariates. The design includes early stopping for futility if no promising treatment effect is identified in the sensitive group and also the difference between the arms in the overall trial population is not significant. The sensitive group is identified using the risk score approach where each patient is assigned a score constructed from their baseline covariates.  
Thus, the adaptive signature allows for the definition of the subgroup during the first stage of the trial, while the enrichment part of the design is based on the subgroup for the second stage. 

We have provided a freely available {\tt R} package that implements the method. We have investigated the performance of the CADEN by applying it to simulated data scenarios with various response rates for the sensitive group and different sample sizes. As expected, the power increased with the increase of the sample size and/or the response rate in the sensitive group due to the fact that the probability of identifying a promising effect in the sensitive group increases. 

We compared the performance of the CADEN design to the non-enrichment CVRS design and showed that the former has higher power in comparison to the latter.   
We also compared the CADEN design to the adaptive enrichment designs where the sensitive group is determined by a single known binary biomarker. Although the CADEN design is inferior to the design where the biomarker is perfectly classified (the benchmark design), it outperforms the other designs when the biomarker is slightly  misspecified which is a more realistic case. The  CADEN design allows one to narrow down the eligibility and also achieves this at a smaller expected sample size, in comparison to the CVRS design. For the null scenario, the CADEN design achieves a well-controlled type I error rate with a substantial reduction in the expected sample size which is a desirable outcome when the sensitive group does not exist. To investigate the behaviour of the design for model misspecification, we analysed scenarios where there exists a subgroup of patients for whom the treatment appears to be harmful. As with the null scenario, the expected sample size is substantially reduced which is again a desirable outcome in these cases. Regarding the power, the CADEN design behaves similarly to the non-enrichment CVRS design.

We investigated different values for the threshold for the significance level of identifying the promising treatment effect in the sensitive group at the interim analysis ($\alpha_2$ = 0.05, 0.1 and 0.2). A larger threshold improves on the power for identifying the treatment effect in the sensitive group. However, when the sensitive group does not exist, increasing the threshold  increases the type I error rate. We suggest incorporating a prior belief regarding the existence of the sensitive group into the consideration for choosing the values of $\alpha_2$, and propose that this issue constitutes a part of further research. 

We illustrated our approach on the randomised clinical trials NOAH and PREVAIL with publicly available high-dimensional gene expression data. However, owing to a relatively small sample size of the trials we sampled an additional set of patients from the trial population. The interim analysis of the NOAH trial resulted in the  ``unselected" strategy because of a significant difference between the treatment arms in the overall trial population. At the final analysis of the NOAH trial, we identified a sensitive group of patients with a significant treatment effect.

In the case of the PREVAIL trial, the interim analysis resulted in the ``stop" strategy (due to a non-significant difference between the arms in the overall trial population and also a non-significant difference between the arms in the in the sensitive group). Here again we treat everyone as belonging to a hypothethical stage one of the trial, due to small sample size. Although the sensitive group does not show a significant difference between the arms, this example illustrates how the data could be further investigated in order to identify a subgroup of patients who benefit from the treatment.

We note that in the case of the ``enrichment" strategy, performing the gene expression profiling would have had to have been done quickly to ensure the smooth running of the trial. Alternatively, there could have been a run-in period prior to the enrolment of patients. An additional caveat would be assessing the impact of the potential misclassification of a genomic classifier that might have lead to a wrong subgroup prediction of patients  \citep{wang:etal:2011}.
In our design, the decision-making is based on the hypothesis testing. In principle, other types of measures could be used for decision-making, such as lower bound of the two sided test, conditional or predictive power, and futility boundaries for early stopping. Another important issue is choosing the necessary sample sizes. Though it has been argued that adaptive enrichment designs require larger sample sizes than standard designs \citep{rosenblum:hanley:2017}, we showed with the real data examples that the new design is applicable for relatively small sample sizes. To increase the power of the design, we maintained the originally pre-planned per-group total sample size \citep{wang:etal:2009}. However, we note, that the power for CADEN will depend on the true sensitive group properties (as if it were a known predictive biomarker) and the sensitivity/specificity that the CVRS method has to classify patients as sensitive. 

We investigated different thresholds for significance level at the interim analysis. Here, a parallel could be drawn to a promising zone design in which the conditional power computed at the interim analysis is classified into pre-specified zones \citep{mehta:pocock:2011}. In practice, we would advise to perform simulations to decide on the appropriate threshold. Another issue to investigate is the timing of the interim analysis. Though the timing depends on the recruitment rate, the systematic review of the promising design showed that the median time to the interim analysis was 60\% of the original target sample size \citep{edwards:etal:2020}. 

To compute the power to reject the dual-composite null hypothesis $H_C$, we split the significance level $\alpha = 0.05$ into two parts: $\alpha_O = 0.04$ (for the power to reject $H_O$), and $\alpha_S = 0.01$ (for the power to reject $H_S$). This split was motivated by the work of \cite{freidlin:simon:2005} and \cite{freidlin:etal:2010} who considered 0.04 and 0.01 for the overall and within-the-subgroup power for the non-enrichment adaptive signature design. Future work will include investigating a number of methods for assigning weights to the statistical significance levels associated with the $H_O$ and $H_S$ hypotheses in order to improve the statistical efficiency of the design \citep{sugitani:etal:2018}. In	clinical practice, the sensitive group might be defined by a certain clinically meaningful threshold $\delta$, i.e. estimated treatment effect $>\delta$. Further research could be focused on investigating alternative clustering approaches that take $\delta$ into account. Additionally, further work could explore different distributions of outcomes, as well as multiple endpoints \citep{cherlin:wason:2021}.

\bibliographystyle{rss}
\bibliography{references}

\end{document}